\documentclass[namedreferences]{solarphysics}
\usepackage[optionalrh]{spr-sola-addons}
\usepackage{epsfig}
\usepackage{graphicx}
\usepackage{color}
\usepackage{url}
\newcommand{\etal}{{\it et al.}}
\usepackage[pdfborder={0 0 0 },urlcolor=blue]{hyperref}
\ifx \doiurl \undefined \def \doiurl#1{\href{http://dx.doi.org/#1}{\url{#1}}}\fi
\ifx \adsurl \undefined \def \adsurl#1{\href{http://adsabs.harvard.edu/abs/#1}{\url{#1}}}\fi

\begin{document}
\begin{article}
\begin{opening}

\title{Will Solar Cycles 25 and 26 Be Weaker than Cycle 24 ?}

\author{J.\ Javaraiah}
\institute{\#58, 5th Cross, Bikasipura (BDA), Bengaluru-560 111,  India.\\
Formerly working at Indian Institute of Astrophysics, Bengaluru-560 034, India.\\
email: \url{jajj55@yahoo.co.in;  jdotjavaraiah@gmail.com}\\
}

\runningauthor{J. Javaraiah}
\runningtitle{Relative amplitudes of upcoming solar cycles}

\begin{abstract}
The study of variations in solar
activity is important for understanding the underlying mechanism
of solar activity and for predicting the level of activity
in view of the activity impact on space weather and global
climate. Here we have used the amplitudes (the peak values of the 13-month
smoothed international sunspot number)  
of Solar Cycles~1\,--\,24 to   
predict the relative amplitudes of the solar cycles during the rising phase of 
the upcoming Gleissberg cycle. 
We fitted a cosine function to  
  the amplitudes and times of the solar cycles  after subtracting a 
linear fit of the amplitudes. 
The best cosine fit shows overall properties (periods, maxima, minima, 
\textit{etc.}) of
 Gleissberg cycles, but with large uncertainties. We obtain a 
 pattern of the rising phase of the upcoming Gleissberg cycle, but 
there is  considerable ambiguity. 
Using  the epochs of violations of
the Gnevyshev-Ohl rule (G-O rule) and the `tentative inverse G-O rule' of
 solar cycles  during the period 1610\,--\,2015,  and also using 
  the epochs where the orbital angular momentum of the Sun
is steeply decreased during the period 1600\,--\,2099,  
  we infer that Solar  Cycle~25 will be weaker than Cycle~24.   
Cycles 25 and 26 will have almost same strength,  
and their epochs are at the minimum between 
the current and  upcoming Gleissberg cycles.  In addition,  
 Cycle~27 is expected to be stronger than Cycle~26 and weaker 
than Cycle~28, and Cycle~29  is expected to be stronger than both 
Cycles~28 and 30.  The maximum of Cycle~29 is expected to 
represent the next Gleissberg maximum. 
 Our analysis also suggests
a much lower value (30\,--\,40) for the maximum amplitude of the
upcoming Cycle~25.
\end{abstract}
\end{opening}


\section{Introduction}

Solar activity varies on many timescales, from a few minutes to a few decades. 
 Solar activity affects us in many ways. It causes problems 
in both the space- and the ground-based technologies, 
impacts on Earth's  magnetosphere and
ionosphere, and also may have an influence on terrestrial climate.
Therefore, predicting solar activity  on all timescales
is important (\opencite{sval13}; \opencite{hath15}).
A number of attempts have been made to predict the amplitude of a sunspot cycle 
by using several  properties of previous cycles~(see \opencite{kane07}; 
\opencite{pes08}, \citeyear{pes12}; \opencite{obrid08}; \opencite{hath09}).
 The amplitude
of a solar cycle near the time of solar cycle minimum can be
predicted  using
precursors such as geomagnetic activity and polar magnetic fields. This method 
seems to  be related to the mechanism 
 of solar cycle, $i.e.$ it is believed that the polar fields act as a 
``seed'' for the dynamo producing the next cycle (\opencite{jiang07} and 
references therein).  From this method 
it seems possible to predict the amplitude of a cycle  
by about four years in  advance with  reasonable accuracy. 
Recently, \inlinecite{camer16}  and \inlinecite{hath16} used simulations of 
the polar field at the minimum of the next  Cycle~25 (at the start of 2020) 
to predict that the amplitude of  Cycle~25  will be similar to that of 
the current Cycle~24. 

\inlinecite{dg06} made an unsuccessful prediction of the amplitude of 
Cycle 24 on the basis of a flux-transport dynamo model. 
 While the solar meridional flow speed varies 
(\opencite{basu03}; \opencite{ju06}; \opencite{jj10}, \opencite{komm15} and references therein),
 in this model a constant value was assumed.
The polar field measurements required as input to the predictions based 
on the flux-transport dynamo models are available only for about three cycles. 
The predictions based on the polar fields at cycle minimum 
 could be  uncertain because 
it was unclear exactly when these polar field measurements 
should be taken. Predictions based on 
the polar fields for previous cycles have given different values at
 different times~(\opencite{hath15}). 
 Sunspot Cycle~24 has two peaks. The second peak is larger than
the first peak (this is unusual) and the gap (about two years) between 
them is also slightly larger than normal. The reason could be that
 Cycle~24 proceeds at different speeds 
in the northern and southern hemispheres.
The southern hemisphere sunspot peak of Cycle~24 occurred in
April 2014 and was about 30\%
stronger than the northern hemisphere peak, which occurred in   
September 2011~(\opencite{hath15}; \opencite{jj16}; \opencite{deng16},
and references therein), 
but the polar 
fields had a similar strength 
in both the northern and southern hemispheres  at the end of Cycle~23.
Therefore, the polar fields may be going 
through different dynamo processes  in the northern and southern hemispheres, 
 which produces two different seeds. That is,  dynamo ingredients seem to  work 
differently in the northern and southern hemispheres, creating 
asymmetries between these hemispheres
(see $e.g.$ \opencite{bd13}; \opencite{shetye15}). Overall, 
 polar fields do not seem to be  a simple precursor.

 \inlinecite{go48} investigated the variability of the sunspot index using the 
Z\"urich database (1755\,--\,1944) and found that the sum of the annual
 international sunspot number ($R_{\rm Z}$)  during 
an odd-numbered cycle exceeds that of the preceding even-numbered cycle, 
 now commonly referred to as the  Gnevyshev-Ohl rule (G-O rule).
 They also showed that 
the sum of annual $R_{\rm Z}$ of an even-numbered cycle correlated well 
(sample size $N = 8$, the correlation coefficient $r = 0.91$ is 
significant on a 99.8\% confidence level)  
with that of the following odd-numbered cycle and that the correlation
was  weak ($r = 0.5$ is significant on 79.3\% confidence level) 
with that of the preceding odd-numbered cycle (also see \opencite{ss97}).
It has been 
believed that it is possible to predict the amplitude of an odd-numbered cycle
using the G-O rule. However, the G-O rule is  occasionally  violated. 
Predictions of  the amplitude of Cycle~23 based on this rule failed 
drastically.   Recently, 
Javaraiah (\citeyear{jj07}, \citeyear{jj08}, \citeyear{jj15}) used 
 the  north-south asymmetry 
property of the sunspot activity (the sum of the areas of sunspot
groups in low latitudes) to show that it is possible to predict
 the amplitude of 
a cycle (valid for both odd- and even-numbered cycles)  
approximately nine years in advance with reasonable accuracy ($N = 11$,
$r= 0.966$ is significant on a 99.99\% confidence level).
 \inlinecite{jj15} predicted that the amplitude of the 
upcoming Cycle~25 will be considerably 
 lower than that of the current Cycle~24, $i.e.$ a violation of the G-O rule
 by the Cycle pair (24, 25).  
 \inlinecite{tltv15} analyzed the numbers of sunspot groups since 1610
  and found that the  G-O rule displays cycles of inversion
with a period of 200 years; the latest inversion occurred in
 Cycle pair (22, 23), and several upcoming 
 odd-numbered cycles should 
be weaker than their respective preceding even-numbered cycles.

\inlinecite{glei39} found a periodicity of seven or eight cycles in
 cycle amplitudes from 1750 to 1928. This property  of the solar 
activity,  known as the Gleissberg cycle, is now  well
established (\opencite{zp14}; \opencite{vaz16}; \opencite{komit16}). 
Some authors have suggested that  the activity is currently at the
minimum of 
the recent Gleissberg cycle (\opencite{jbu05a}; \opencite{zp14}; 
\opencite{gao16}). In this article  
we attempt to predict the relative amplitudes of the solar cycles that 
correspond to  the 
rising phase of the upcoming Gleissberg cycle. For this, we have 
analyzed the  yearly and the cycle-to-cycle  variations in $R_{\rm Z}$.
 Essentially, we  use the  aforesaid prediction 
 by \inlinecite{jj15} and   long-term trends in the yearly and 
cycle-to-cycle variations in  $R_{\rm Z}$. 
We  also use the epochs of the steep decrease in the
 Sun's orbital angular momentum about the solar system barycenter, 
which seem to help inform  epochs of violations of G-O rule 
(\opencite{jj05}). That is, we  use these epochs 
  to identify some long-term trends in 
$R_{\rm Z}$ for the present purpose. Our aim here is not
 for to establish the solar activity connection to the solar system dynamics 
(which is  beyond the  scope of the present analysis). 

In the next section we describe the data analysis and results.  
  In Section~3 we summarize  the conclusions and present 
a  brief discussion.

\begin{figure}
\centerline{\includegraphics[width=\textwidth]{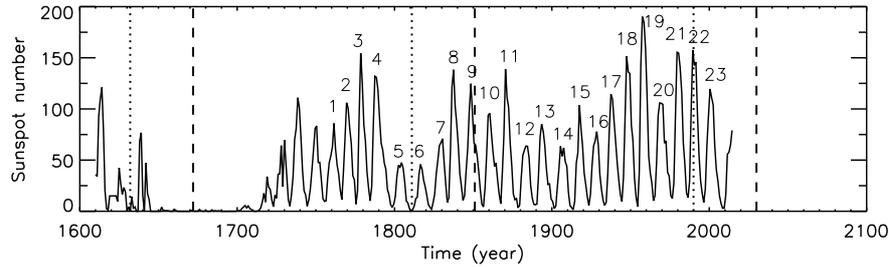}}
\caption{Variations in the yearly mean $R_{\rm Z}$ (before 
1950, the data represent the observations that were compiled by Hoyt and Schatten (1998) from numerous sources).
Near the peak  of each solar cycle the corresponding
Waldmeier cycle number is given. The {\it dotted vertical lines}
(in 1632, 1811, and 1990)
and {\it dashed vertical lines} (in 1672, 1851, and 2030) 
are drawn at the epochs of steep decrease in  the Sun's orbital 
angular momentum about the solar system barycenter. Close to all these 
 epochs, the orbital motion of the Sun is retrograde (the  rate 
of change  in orbital angular momentum of the Sun is slightly negative).
At the epochs that are indicated with {\it dotted vertical lines},
the values of the Sun's
distance from the solar system barycenter, orbital velocity,
 and orbital angular momentum are slightly higher than those at 
the epochs that are indicated with {\it dashed vertical lines}
 (see Table~1 in Javaraiah, 2005).}  
\label{Fig.1}
\end{figure}

\begin{figure}
\centering
\includegraphics[width=\textwidth]{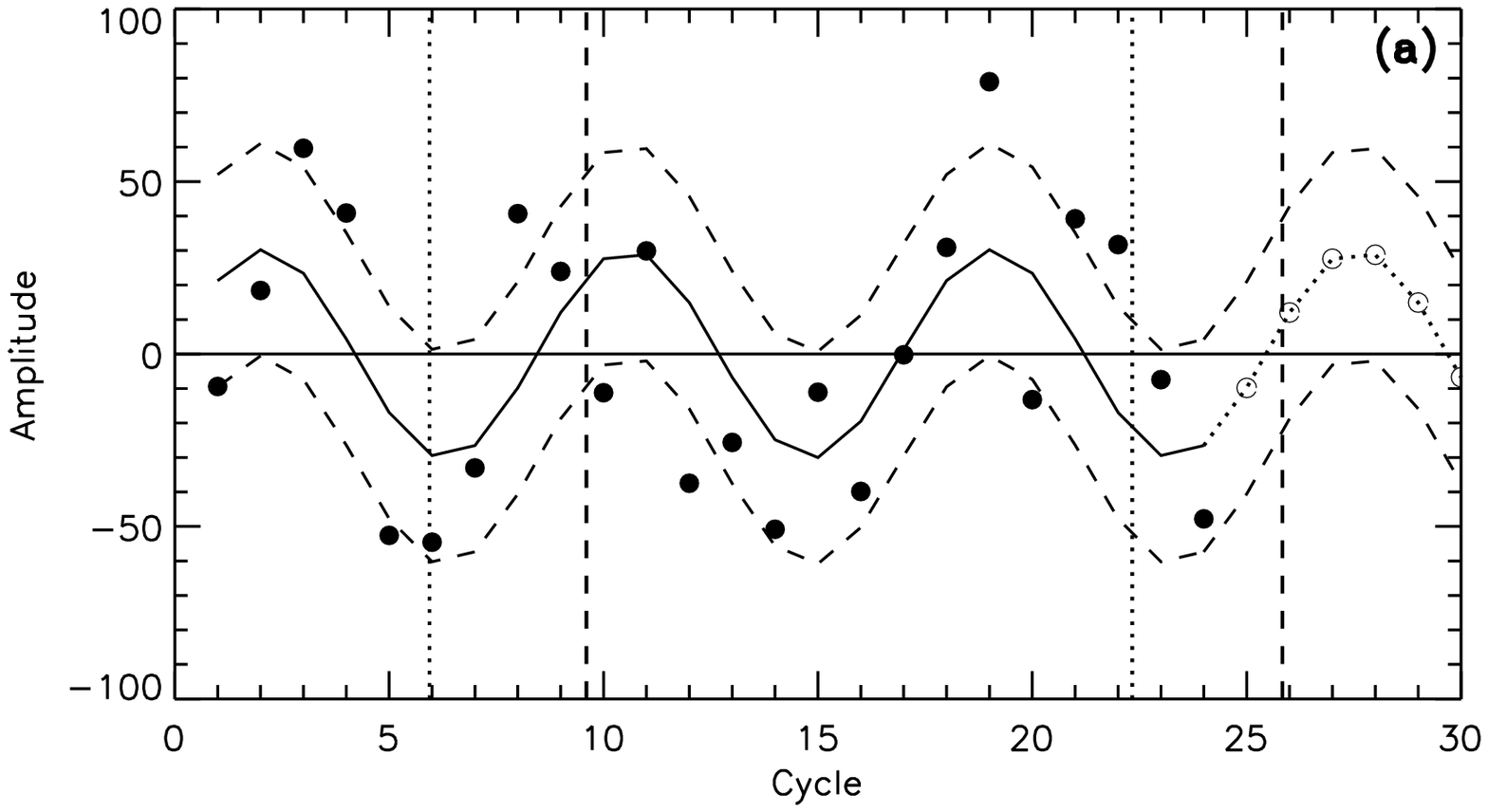}
\includegraphics[width=\textwidth]{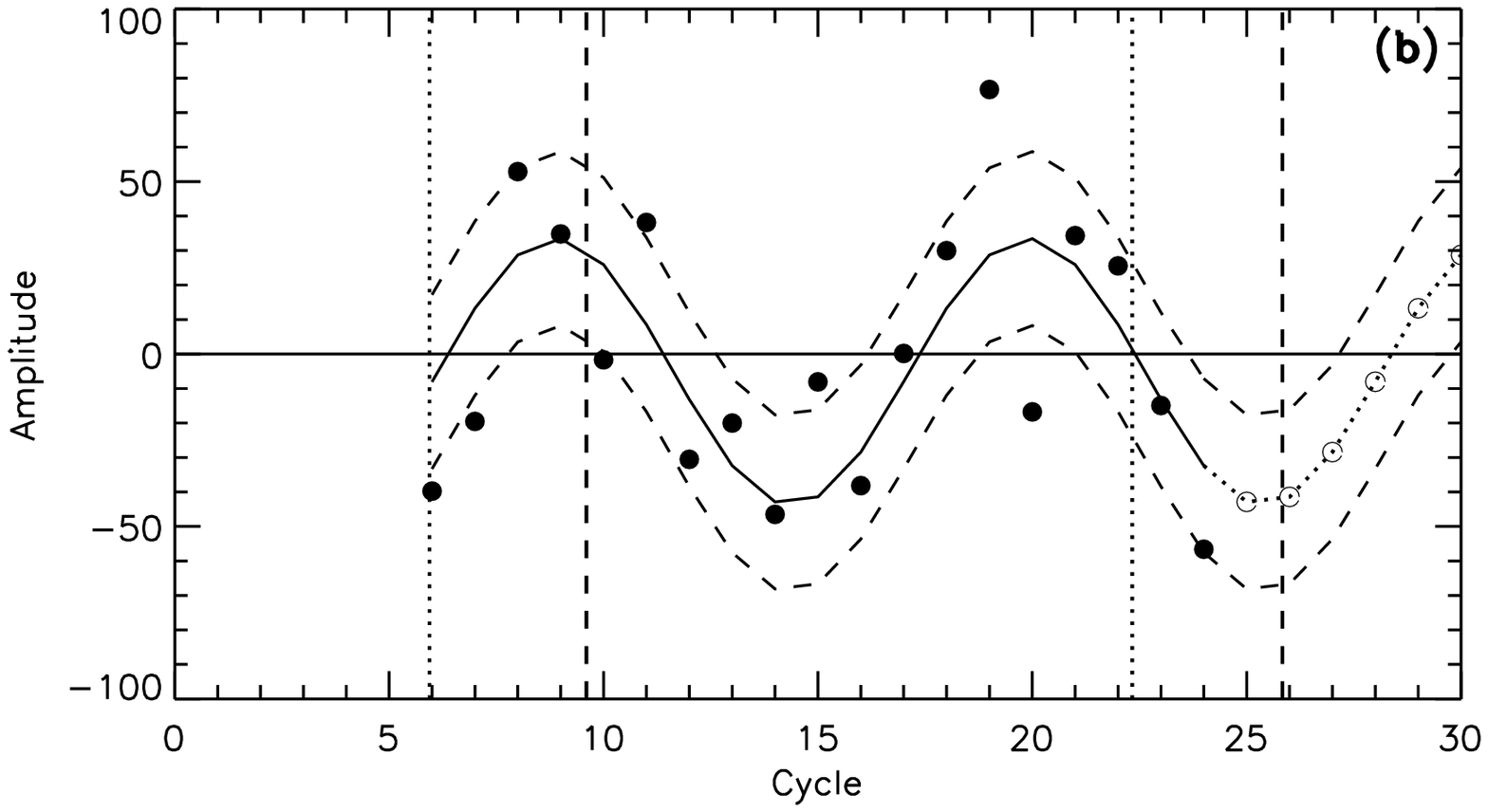}
\caption{Plot of the
amplitudes ({\it filled circles}), $i.e.$ the peak values of the 13-month 
smoothed $R_{\rm Z}$, of the sunspot cycles after  subtracting a linear fit  
from the cycle amplitudes as a function of time.
 The {\it continuous curve} represents the best  cosine fit.
The {\it open circles} connected by the {\it dotted curve} represent the
 extrapolated  amplitudes of the upcoming Cycles 25\,--\,30.   
The {\it dashed curve} represents the {\it rms deviation} ($\sigma$). 
Same as Figure~1: the {\it dotted} and {\it dashed vertical lines}  
are drawn at the epochs of steep decrease in  the Sun's orbital 
angular momentum. 
(a)  The entire cycle data set (1\,--\,24), the corresponding  $\sigma = 30.8$ and $\chi^2 = 433$, and 
(b)  Cycles 6\,--\,24 data, the corresponding $\sigma =25.2$ and
 $\chi^2 = 237$.}
\label{Fig.2}
\end{figure}

\section{Data analysis and results}
Figure~1 shows the variation in the yearly mean 
 $R_{\rm Z}$ during the period 
1610\,--\,2015, which is taken from {\tt http://www.ngdc.noaa.gov/}.
 The same figure also shows the epochs 
 when the orbital angular 
momentum of the Sun about the solar system barycenter decreased steeply
 (about 25\% lower than mean  
drop; also see  \opencite{jj05}) during the period 1600\,--\,2099.
Close to  these
 epochs the orbital motion of the Sun is retrograde, $i.e.$ the  rate
of change in  orbital angular momentum of the Sun is slightly negative.
However, at these epochs
 the orbital positions of the giant planets are different, hence
there are slight differences in the corresponding values of the Sun's
distance from the solar system barycenter, orbital velocity, orbital angular
momentum and its rate of change (see Table~1 in \opencite{jj05}).  
 In 1632, 1811, and 1990 (indicated with dotted vertical lines), 
the values of the Sun's 
distance from the solar system barycenter, orbital velocity,
 and orbital angular momentum are slightly higher than those in 1672, 1851, 
and 2030 (indicated with dashed vertical lines). 
 The orbital angular momentum  data  were provided by  Ferenc V\'aradi.  
He had derived these data using the  Jet
Propulsion Laboratory (JPL) DE405 
ephemeris for the period 1600\,--\,2099.
The positions and velocities required may be obtained from JPL's online 
Horizons ephemeris system 
(see \opencite{seid92}; \opencite{giorgini96}; \opencite{stan98}).
We have fitted a cosine function to the amplitudes of the solar cycles
(the peak values of  13-month smoothed  $R_{\rm Z}$, which are also 
taken from {\tt http://www.ngdc.noaa.gov/})   after removing the 
long-term trend by subtracting a linear fit to
the amplitude values as a function of time, 
in a similar manner as  was  done  
for the strengths of Cycles 6\,--\,23 by \inlinecite{jbu05a}. 
We  determined the best-fit cosine function for all 
Cycles 1\,--\,24  and only for Cycles~6\,--\,24. 
The results 
are shown in Figures~2(a) and 2(b). 
In these figures the extrapolated values for Cycles 25\,--\,30 are also
shown. The fitting of the Cycles 6\,--\,24 data
 ({\it root-mean-square (rms) deviation} $\sigma = 25.2$ and  
$\chi^2 = 237$)  is slightly  better 
than that  of all Cycles 1\,--\,24 data ($\sigma =30.8$ and $\chi^2 = 433$).
 The values of $\chi^2$ are very high 
(the values $\chi^2$ significant at 17 and 23 degrees of freedom are
 27.587 and 35.172, respectively,   at the
5\% level of significance, $i.e.$ for $P = 0.05$). However,  
 the difference between the $\chi^2$ values 
  are not critical for the current study (see below).

The continuous curves in both  Figures~2(a) and 2(b) represent
the Gleissberg cycles, but have a substantial 
difference in  their periods: 
the former shows the cycles of  period  8.5 solar cycles, 
  whereas the latter shows the cycles of period  11.0 solar cycles. 
 Several authors have found  changes in  the period of Gleissberg 
cycle~(\opencite{garc98}; \opencite{hath99}; \opencite{roze94}; 
\opencite{ogurt02}). 
\inlinecite{hath15} examined the amplitudes 
of Cycles 1\,--\,23 and found a 9.1-cycle periodicity.
In both  Figures~2(a) and 2(b) the minima of the Gleissberg 
cycles are reasonably well defined (lower uncertainty, i.e. the data points 
are within the dashed curve drawn at 1$\sigma$ level). 
However, Figure~2(b) 
clearly shows that  Cycles~25 and 26 are at the minimum  between the current 
and the upcoming  Gleissberg cycles, whereas Figure~2(a) shows that the 
corresponding minimum is at Cycle~24. That is, there is a considerable 
difference between  the extrapolated portions of the  
curves in Figures~2(a) and 2(b), which  show different epochs for 
 the minimum and the maximum of the upcoming Gleissberg cycle. 
Overall  the   uncertainties
in the cosine fittings  are  large ($\chi^2$  values are much higher  
than the corresponding values at the 5\% level of significance).   
Hence,  it is not possible to infer the  amplitudes of the 
upcoming cycles or even their relative values from either of 
the curves shown in Figures~2(a) and 2(b). In each of these figures
the extrapolated values have large uncertainties.
From the trends in $R_{\rm Z}$ shown in Figure~1
may be   possible to  infer  
the  relative values of the amplitudes of the upcoming solar cycles 
as argued below.

As can be seen in Figure~1,  the Cycle pairs (4, 5), 
(8, 9), and (22, 23) violated the  G-O rule. Excluding these cycle pairs
 we find  $r = 0.972$ (significant on the 99.99\% confidence level),
 the slope is 10.6 times larger than the standard deviation, and
$\chi^2 =6.43$ (insignificant on 5\% level) of the best-linear fit
between amplitudes of the remaining  pairs ($N = 8$) of even- and the following
odd-numbered cycles (see Figure~3(a)). 
Although, as mentioned in Section~1, the correlation between the sizes
of an even-numbered cycle and  preceding odd-numbered cycle  is weak,  
 Figure~1 shows that  the amplitude of 
an even-numbered cycle is frequently  smaller than its preceding 
odd-numbered cycle. 
(In the case of Cycle pairs (5, 6) and (21, 22) the amplitudes of 
the odd-numbered Cycles 5 and 21 are negligibly larger than those of the 
even-numbered Cycles
6 and 22) That is, there is a qualitative relation between
an even-numbered cycle and its preceding odd numbered cycle.
This property,  which may be called the `tentative inverse G-O rule', of
  even-numbered  
and preceding odd-numbered  cycles,    is
violated by  Cycle pairs (1, 2), (7, 8), and (17, 18). 
 Excluding these cycle pairs, we find 
 $r = 0.77$ (significant on 98.48\% confidence level), 
the slope is 3.3 times larger than the standard deviation,  and
$\chi^2 =36.5$ (highly significant on 5\% level)  of the best-linear fit
between amplitudes of the remaining  pairs ($N = 9$) of odd- and the
 following even-numbered cycles (see Figure~3(b)). 
 Overall, this relationship has a large uncertainty.
In the case of small sunspot groups (maximum area smaller than 100 MSH), 
the inverse G-O rule found to be valid throughout  
Cycles 12\,--\,24 (\opencite{jj12}). Using the relationships shown 
in Figure~3
and by knowing the epochs of  violations of the G-O rule and the
 inverse G-O rule,  
it may be possible to predict the amplitude of a
cycle from these rules (prediction for an even-numbered cycle is qualitative)
except for the odd- and even-numbered cycles  of the cycle pairs that  
violate the G-O rule and  the inverse G-O rule.
It may be noted here that the G-O rule  and 
its violation has been known for a long time (see \opencite{komit01}).  
The inverse G-O rule and its violations suggested above  are speculative. 

\begin{figure}
\centering
\includegraphics[width=9cm]{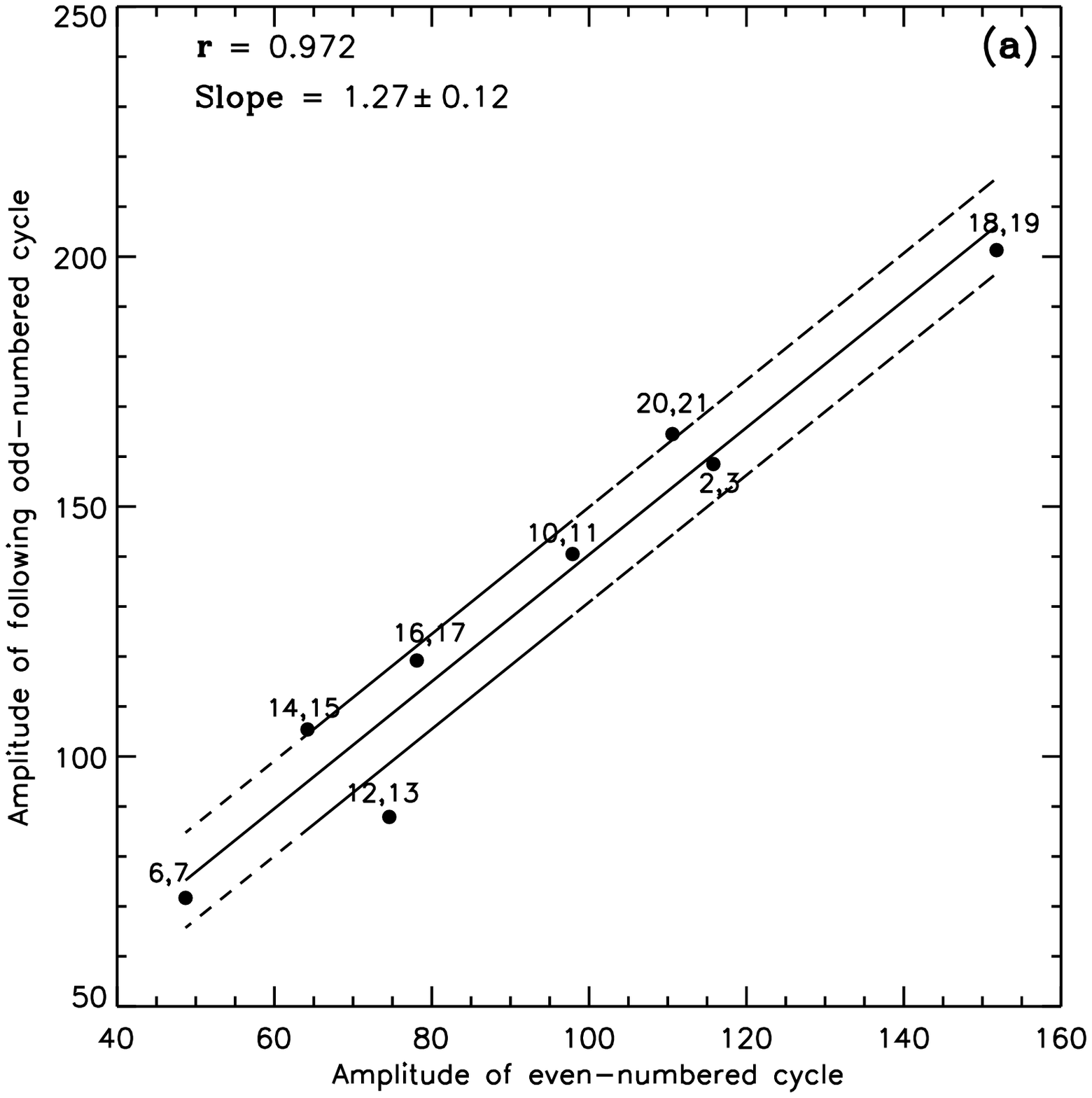}
\includegraphics[width=9cm]{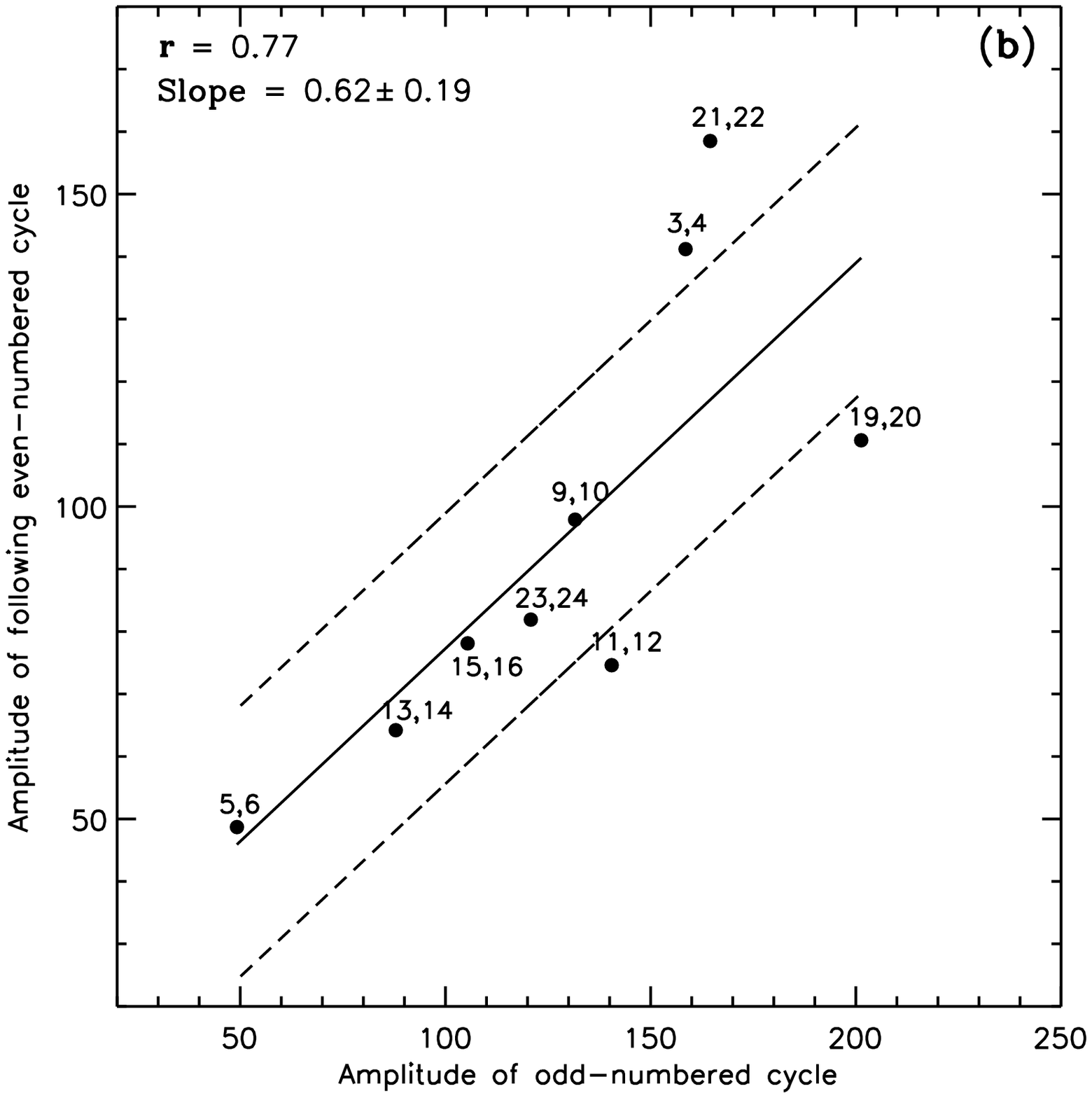}
\caption{Plots of  amplitudes:  (a) even-numbered cycles versus 
 following odd-numbered cycles that satisfy the G-O rule and (b) odd-numbered 
cycles versus following even-numbered cycles that satisfy the tentative 
inverse G-O rule. The {\it continuous curve} represents the 
best linear fit and the  {\it dashed curve} represents {\it rms deviation} 
($\sigma$). The values of the correlation coefficient ($r$)  and of the
 slope are also given, and near each data point, the corresponding cycle
 pair is shown.} 
\label{Fig.3}
\end{figure}

As can be seen in Figure~1,  the epochs of the Cycle pairs (4, 5), 
(8, 9), and (22, 23) that violated the  G-O rule
 are close to the epochs when the Sun's orbital 
angular momentum is steeply decreased. Since it is possible to know
  the future epochs of the steep decrease in  the Sun's orbital
 angular momentum well in advance,  it seems possible to predict 
the violation of the G-O rule well 
in advance as well~(\opencite{jj05}).    
The orbital angular momentum of the Sun is steeply decreased 
during  alternating  
 40-year and 139-year intervals. The next such a situation will 
arise in 2030 and after that, in 2169. Therefore,   
the next   violation of the G-O rule is expected to occur 
in  2030, and after that,   in 2169. 
 The amplitude of Cycle~25  predicted to be lower 
than that of Cycle~24~(\opencite{jj15}). 
 Since  small cycles usually have long 
periods, the upcoming  Cycle~25 may also be a long cycle. 
We determined  linear fit to  the values of  amplitudes and 
lengths of Cycles 1\,--\,23.
 The values of lengths (the time intervals
between consecutive activity minima) were also taken from 
 {\tt http://www.ngdc.noaa.gov/}. 
 ($N = 23$, $r = -0.35$ is small,
 also see \opencite{solanki02}.  
The corresponding value of the 5\% level significance  is 
about -0.41 for 21 degrees of freedom.) 
Using the value 81.9  of the amplitude  of Cycle 24  and the 
values  30\,--\,50 predicted 
for Cycle~25 (see \opencite{jj05} and below) 
in the best-fit linear equation, we obtain   
the values 11.4 and 11.7\,--\,11.9 for the 
 lengths of Solar Cycles 24 and 25, respectively. 
Therefore, Cycle~24, which 
 began in 2008.9, will probably end close to 2020, and Cycle 25 may
 end close to 2032 (these estimates have large uncertainties).   
These  estimates suggest that  the upcoming Cycle 25
 will include the epoch 2030, when the  next steep decrease in the 
 orbital angular momentum of the Sun will take place.
  This cycle will be at the minimum
 between the current and the next Gleissberg cycles (see Figure~2).

The average level of activity is drastically different for
 different occasions 
of the steep decrease in the orbital angular momentum of the Sun
(see Figure~1). 
Both the Maunder minimum (1645\,--\.1715) and  Dalton 
minimum (1790\,--\,1830) included an epoch of the steep decrease in the
orbital angular momentum of the Sun.
However, in  1851 and 1990, when  the orbital angular momentum of 
the Sun  steeply decreased the level of activity was reasonably high.
Between the dotted and dashed  vertical lines 
 in the intervals 1632\,--\,1672, 1811\,--\,1851, and 
1990\,--\,2030, the activity  decreased, increased, and decreased,
respectively. The decreasing trend during
1990\,--\,2030 may lead to a considerably low activity around 2030, 
but it may be significantly higher than that in the Maunder minimum.
 A best linear fit to the values of amplitudes 
(164.5, 158.5, 120.8, 81.9, 30\,--\,50)  and times 
of Cycles 21\,--\,25 (predicted values are used for Cycle~25)
 suggests 7\,--\,23 for the amplitude of Cycle~26, which 
is expected to begin close to the  epoch 2030.
Planetary configurations 
may cause modulations in the emerging magnetic flux  
 (\opencite{jj05}; \opencite{jb16}). 
It should be noted that no alignment of the major planets is exactly repeated
(\opencite{jose65}).  

 Most of the solar system angular momentum is contributed by the 
 orbital motions of the four giant planets, whose time-dependent 
spatial configurations are responsible for the irregular orbital motion of the 
Sun around the solar system barycenter.
 The orbital motion of Jupiter is the primary contribution (about 60\%),  
and the Sun's spin  
 contributes about one to two percent to the total angular
 momentum of the solar system. (The net external torque on the Sun, 
due to orbital motions of the planets,  is not
 zero, although it is small. So there could be some variation in the 
spin angular momentum of the Sun.) 
The solar rotation varies on many timescales 
(see \opencite{seh85}; \opencite{snod92}; \opencite{kj02}; \opencite{antia02};
\opencite{jbu05a}, 2005b; \opencite{braj06}; \opencite{jj13}; \opencite{jb16}).
\inlinecite{jj05} foun that significant 
(at the 99.5\% confidence level) positive (before around 1945) and negative 
(after around 1945)
correlations exist between the Sun's orbital torque and 
the solar equatorial rotation rate determined
 from sunspot group data during the period 1879\,--\,2004.
 Differential rotation and meridional flow are both important
 ingredients in flux-transport dynamo models (\opencite{dg06}).
Meridional circulation is primarily determined by the Coriolis forces
 from differential rotation, turbulent Reynolds stresses, and pressure
and buoyancy forces  (\opencite{rud89}). An approximate 
linear relationship exists between the solar cycle variations in differential 
rotation and the meridional velocity of sunspot groups (\opencite{ju06}).
 A high correlation (significant at the 99\% confidence level) 
exists between the
  meridional flow and  the Sun's orbital 
torque during Cycle~23 (personal communication from J. H. Shirley in 2017). 
(J.H. Shirley 
analyzed high-resolution JPL Horizons 
ephemeris data to determine the Sun's  orbital torque  and  
used the meridional flow speeds of small 
 magnetic features on the Sun determined by \inlinecite{hr10}  from  
 observations of the \textit{Michelson Doppler Imager} (MDI) on board the
\textit{Solar and Heliospheric Observatory} (SOHO) spacecraft spanning a
 period from May 1996
 to June 2009. The values of the orbital torque and  meridional flow speeds
  were averaged over time periods corresponding to one Carrington rotation,
and the resulting 167-data-point time-series were cross correlated.)
  The variations in the Sun's spin-momentum may relate
  to variations in the solar 
differential rotation and therefore the solar 
dynamo (\opencite{zaq97}; \opencite{juc03}; \opencite{wil08}). 
On the other hand, the Coriolis force due to solar rotation affects  the 
rising magnetic flux through the convection zone to the photosphere due 
to magnetic buoyancy (\opencite{chg87}). 
Variations in  solar rotation may be related to 
the variations in solar activity through the effect of Coriolis forces
 on the emerging solar magnetic flux.

The  values of the Sun's orbit 
 and spin angular momenta are 
$\sim 5\times 10^{47}$ g cm$^2$ s$^{-1}$ (maximum) and $\sim 2\times 10^{48}$ g cm$^2$ s$^{-1}$
(\opencite{antia02}), respectively. 
 \inlinecite{jj05} found that   a good agreement  exists between 
the  amplitudes of the variations in the Sun's spin and the orbital 
angular momenta, particularly at the common epochs of the steep decreases
 in  both the Sun's orbital angular momentum and the equatorial 
rotation rate determined from sunspot data.  
 At these epochs the orbital angular momentum is low an order of magnitude
 approximately equal to  its own magnitude (the magnitude varies
from 0 to $\sim 10^{47}$ g cm$^2$ s$^{-1}$). For example, 
at  the epoch 1990.97,  it is found to be $-2.1\times 10^{47}$
 g cm$^2$ s$^{-1}$, and there  
the amount of the drop  in the equatorial 
rotation rate is found to be  about 1\% and the corresponding 
spin momentum is approximately equal to   $-1.1\times 10^{47}$
 g cm$^2$ s$^{-1}$.
 The epochs of violations of the G-O rule are 
 close to the common epochs of the steep decrease in the orbital angular 
momentum of the Sun (when the rate of its change is slightly negative, 
 $i.e.$ when the  orbital motion of the Sun is retrograde)
 and the equatorial rotation rate (see Figure~2 in \opencite{jj05}).
 A correlation exists between the  
 equatorial rotation rate  and the latitude gradient of rotation 
 (\opencite{balth86}). Hence,
a decrease in the equatorial rotation rate  also implies  
a decrease in the differential rotation and  leads to
 a weak dynamo. 
A few weak cycles follow a large  
drop in the equatorial rotation rate (\opencite{jj03a}). (Note that 
in the  equatorial rotation rate  a considerable drop occurred 
  from Cycle~13 to Cycle~14 (see \opencite{balth86};  
\opencite{jj03a}, 2003b; \opencite{jbu05a}), $i.e.$ the Gleissberg cycle
 minimum around Cycle~14 
also followed a considerable drop in  equatorial rotation rate.)
\inlinecite{rr93}
found that during the  Maunder minimum, the equatorial rotation rate
was lower by about 2\%. \inlinecite{vsg02} found that it was lower
 by about 5\%. (However,  many studies show  
 that  rotation rate,   equatorial rotation rate, and  
  latitude gradient are higher  during a solar cycle minimum than during  
 maximum ($e.g.$, \opencite{balth86}; \opencite{khh93}; \opencite{jk99}; 
 \opencite{jj03b}; \opencite{braj06}).
The large drops in the long-term  rotation variation  may originate  at
 deeper layers of the Sun,
 while the high rotation rate  during a solar cycle minimum may represent 
 the rotation rate of small magnetic regions at subsurface  and surface layers, 
see \opencite{jj13}; \opencite{jb16} and references therein).

The violation of G-O rule is followed by at least one  relatively weak cycle
(see Figure~1). Since Cycles 6, 10, and 24 are weaker than Cycles 5, 9,
and 23, respectively, Cycle~26 will be a weak cycle whose strength  may be 
close to that of Cycle 25 (note that as argued above the Cycle pair (24, 25) 
is expected to violate the G-O rule,  and also note the
prediction of  \inlinecite{jj15}).
 \inlinecite{komit01} have found that violation of the G-O rule 
cannot be a random phenomena, but occurs
under special conditions, the main factor being the very high
maximum of the even-numbered  cycle, for example, Cycles 4, 8, and 22. 
Since  the maximum of Cycle~24
 is not very high,  it does not support the notion that the Cycle pair (24, 25) 
 will violate the G-O rule as  predicted by \inlinecite{jj15}.   

 The basic concept of the   prediction  of \inlinecite{jj15} is 
 the existence of a high correlation ($r$ is 0.96\,--\,0.97)  between the 
sum of the areas  of sunspot groups in the $0^\circ - 10^\circ$ 
interval of the southern 
hemisphere during the time interval of 1.0 year to 1.75 
year just after the 
time of the maximum of a cycle and the amplitude   of its 
immediately following cycle (\opencite{jj07}). A statistically
high significant linear relationship exits between these quantities 
($cf.$ Equation~(2) in \opencite{jj07}; Equation~(3) in \opencite{jj15}).   
In the case of Cycle~24, the second  peak of $R_{\rm Z}$  is larger than 
the first peak, which is unusual. The first peak with a value of
 66.9  occurred in 
2012.17, while the large second peak with a value of 81.9 is occurred
 in 2014.3.
\inlinecite{jj15} analyzed the combined Greenwich and
 \textit{Solar Optical Observing Network} (SOON)
 sunspot group daily data during the period 1874\,--\,2013 
(downloaded from the website
 {\tt http://solarcience.msfc.nasa.gov/greenwich.shtml}) and 
determined the sums of the areas of sunspot groups in the different 
$10^\circ$ latitude intervals of the northern and southern hemispheres
 during different time intervals.
 By using the value 18.86 of the sum (divided by 1000) of the 
areas  of sunspot groups in $0^\circ -10^\circ$ latitude interval of the 
southern 
hemisphere during  the interval 2013.17\,--\,2013.92, 
which is one year away from the epoch of the first peak of Cycle~24, 
 \inlinecite{jj15}
 predicted $50 \pm 10$ for the amplitude 
  of the upcoming solar Cycle~25.
 Here we analyzed the extended Greenwich-SOON data (extended with
 the SOON sunspot group daily data for the period 2014\,--\,2016)  and 
 determined the sum of the areas of sunspot groups 
 in the $0^\circ - 10^\circ$ latitude  interval of the southern hemisphere 
 during the interval 2015.30\,--\,2016.05 
which is one year away from the epoch of the large second peak of Cycle~24. 
This sum is found to be only 5.18. 
Hence, we obtain a much lower value $29.9 \pm 10$ for the amplitude of
 Cycle~25. 
 Here we have also analyzed the combined updated Greenwich and 
Debarcen Observatory sunspot group daily data (DPD)  during 
the period 1874\,--\,2016, which were downloaded 
from the website {\tt http://fenyi.solarobs.unideb.hu/pub/DPD/} 
(for details see  \opencite{gyr10}).
 We find a relation 
 that is same as the 
Equation~(3) in \inlinecite{jj15}.
The sums of the areas of sunspot groups 
in the $0^\circ - 10^\circ$ latitude interval of the 
southern hemisphere during the  time
 intervals 2013.17\,--\,2013.92 and 2015.30\,--\,2016.05
are found to be 28.19 and 11.0,  respectively.  Using these 
values,  we derive the values  
$65 \pm 11$ and  $39 \pm 11$ for the amplitude of Cycle~25. 
As we have pointed out earlier 
(\opencite{jj07}, \citeyear{jj08}, \citeyear{jj15}), the interval of 
1.0\,--1.75 year after a cycle maximum epoch is close to (or includes) 
the epoch of the reversal of the Sun's polar magnetic
 field (see Table~1 in \opencite{mts03}).
 The magnetic field in the southern
 hemisphere reversed its polarity in mid-2013 (\opencite{mord16}), 
$i.e.$ within the 
intervals 2013.17\,--\,2013.92, which is one year away from 
the epoch of the first peak of Cycle~24 (there may be some transport of 
magnetic flux across the solar equator,  the combined effect of the 
Sun's rotation and the inclination of the Sun's equator may play a role). 
The   values,  which were obtained for the amplitude of Cycle~25  using the
sums of the areas of the sunspot groups in this interval, seem to be fine.
 However, the low values (30\,--\,40) predicted 
 using the  sums of the areas of sunspot groups 
 in the interval 2015.30\,--\,2016.05, which is  one year away from  
the  epoch of the large second peak of Cycle~24,  are more relevant to the
 basic concept described above. These values imply that Cycle~25 will be a
 much weaker cycle.     

 The  values $74 \pm 10$ (\opencite{jj05}) and
 $87 \pm 7$ (\opencite{jj08}) which were predicted for the amplitude
 of Cycle~24 
are close to the real value, $i.e.$ these values match
 (within their uncertainty limits)  the   
value 81.9 of the maximum  amplitude of Cycle~24
 (all the other  predictions in the two articles cited last have lower
 uncertainties, and all indicated that Cycle~24 is weaker than  Cycle~23).  
Hence, mostly  the amplitude of Cycle~25 will be lower than 
that of Cycle~24, as predicted in \inlinecite{jj15} and here.
On the other hand, although we here predict  much lower values  
 for the amplitude of Cycle~25 by  analyzing    
two data sets, assuming that the prediction fails, $i.e.$ the
 Cycle pair (24, 25) will not violate the G-O rule, hence,
   the property that  violation of 
G-O rule is followed by at least one weak cycle will  not be applicable 
for Cycle~26. However,
 Cycle~26 is still expected to be weaker than 
Cycle~25. This is because an epoch of the steep decrease in the orbital
 angular momentum of the Sun seems to be followed by a relatively 
weak cycle,  and  Cycle~26 will follow the steep decrease in the 
orbital angular momentum of the Sun at 2030.
 This is also consistent with the 
tentative inverse G-O rule, as well as  with the  decreasing 
trend of the activity from Cycle~22, $i.e.$, from the last dotted vertical 
line in Figure~1, as suggested above. Even if the Cycle pair (25, 26)
  will not satisfy the inverse G-O rule,   the  amplitude of 
Cycle~26 will still not be much higher (see the trend in the extrapolated
 portion of the cosine curve in Figure~2(b)).

The violation of the tentative inverse rule also seems 
to occur in somewhat regular intervals. As we described above, 
 the Cycle pairs (1, 2), (7, 8), and (17, 18) 
violated this tentative rule. The first of these pairs is in the fourth 
quarter of the 139-year interval of the 
orbital angular momentum of the Sun, which comprises
 part of the Maunder minimum, the second one is in the 40-year interval, 
 which comprises part of the Dalton minimum, and the third one is in the
fourth quarter of the  139-year interval, which comprises
the modern minimum (1890\,--\,1939).  This sequence shows that   
the next one should take place in 
the 40-year interval that
comprises the current deep minimum. However, the difference between the 
epochs of the first two pairs that  violated the tentative inverse G-O rule 
is four cycles (3\,--\,6), while the differences between the second and 
third pairs  is eight cycles (9\,--\,16). Already more than five cycles
 elapsed since the last Cycle pair (17, 18) that  
violated the tentative inverse G-O rule,  
 and as inferred above, Cycle pair (25, 26)
is expected to  obey the 
tentative inverse G-O rule. Thus, if we consider the difference of 
eight cycles, 
the next cycle pair that will  violate the tentative inverse G-O rule 
will be cycle pair (27, 28). Hence, Cycle~27 will be weaker than Cycle~28. 
In addition, Cycles 1, 7, and 17 are 
stronger than their respective preceding even-numbered cycles. Hence,  
we can also infer that Cycle~27 will be stronger 
than Cycle~26. As described above, the next epoch of the violation of 
 G-O rule is 
much farther away, hence Cycle~29 will be  stronger than Cycle~28.
Even when we consider the difference of four cycles,  after
 Cycle pair (27, 28), the next cycle pair that is expected to  
violate the tentative inverse G-O rule may be cycle pair (33, 34).
Hence, Cycle~29 is also expected to be stronger than Cycle~30.  
Cycle~29 will probably  be  a substantially strong cycle,
 and its  maximum may represent the upcoming Gleissberg maximum. 
However, we cannot rule out 
the possibility that the maximum of  Cycle~31 represents  the upcoming 
Gleissberg maximum, as we can estimate from the  curve shown 
in  Figure~2(b).  Overall, the inferences above are consistent with the
 pattern of the extrapolated portion of the curve in Figure~2(b),  $i.e.$ 
the values of the amplitudes  of Solar Cycles 25\,--\,30 are expected to be
 close to the corresponding extrapolated values and satisfy 
the aforesaid inferences and predictions (listed in the next section)  
 on the relative amplitudes of these solar cycles.

\section{Conclusions and discussion}
The solar activity varies on many timescales.
The study of variations in solar
activity is important for understanding the underlying mechanism
of solar activity and for predicting the level of activity
in view of the activity impact on space weather and global
climate. Here we have used the 
amplitudes of Sunspot Cycles~1\,--\,24 to   
predict relative amplitudes of the solar cycles during the rising phase of 
the upcoming Gleissberg cycle. 
We fitted a cosine function to  
  the amplitudes  and times of solar cycles,  after subtracting a 
linear fit of the amplitudes. 
The best cosine fit shows overall properties (periods, maxima, minima,
 \textit{etc.}) of
 Gleissberg cycles, but with large uncertainties. We obtained a 
 pattern of the rising phase of the upcoming Gleissberg cycle, but 
there is considerable ambiguity. 
Using  the epochs of violations of
the Gnevyshev-Ohl rule (G-O rule) and the `tentative inverse G-O rule' of
 solar cycles  during the period 1610\,--\,2015,  and also using 
  the epochs where the orbital angular momentum of the Sun
is steeply decreased during the period 1600\,--\,2099,  
  we inferred that 
i) Solar  Cycle 25 will be weaker than Cycle 24, ii) 
Cycles 25 and 26 will have almost same strength,
and the epochs of these cycles will be at the minimum between
the current and  upcoming Gleissberg cycles,
iii) Cycle 27 is expected to be stronger than Cycle~26 and weaker
than Cycle~28, iv) Cycle 29  is expected to be stronger than both
Cycles~28 and 30, and v) the maximum of Cycle~29 is expected to 
represent the next Gleissberg maximum.  Our analysis also suggests 
a much lower value (30\,--\,40) for the maximum amplitude of the 
upcoming Cycle~25.

 It should be noted that  the   physical connection between the solar 
system dynamics and solar activity is not clear yet
(\opencite{zaq97}; \opencite{juc03}; \opencite{mhg10}; \opencite{wolff10}; 
\opencite{abreu12}; \opencite{cc12}; \opencite{wil08}; \opencite{wil13};
  \opencite{chow16};  \opencite{stef16}). However, 
the aim of the present analysis is not to establish this connection.
The relatively short-term
predictions of the solar activity that were made based on the
hypothesis of a role of solar system dynamics in the mechanism
of solar activity have failed (\opencite{mees91}; \opencite{li01}).
 \inlinecite{char08} used the   
motion of the Sun about the barycenter to infer  that
 Cycles 24\,--\,26 will be repeat
  Cycles~11\,--\,13, and they 
predicted 140 (also 100), 65 and 85  for the amplitude of  Cycles~24, 25 and 26,
 respectively. Obviously, the prediction for Solar Cycle~24 failed, 
but the trend over Cycles~24\,--\,26  is consistent
 with our prediction that both Solar Cycles 25 and 26
 will be weaker than Cycle~24 (conclusions (i) and (ii) above).
 However, only the hypothesis of a role of solar system dynamics in 
the mechanism of solar activity is not a  main factor in the
 present analysis.
 In fact, only one necessary piece of information, namely the epochs of the
 steep decreases 
in the orbital angular momentum of the Sun, was  used for  our predictions  
above  (conclusions (i)\,--\,(v) above).
According to these epochs,  the violation of the G-O rule is much
 farther way after 
the violation in  2030 (\opencite{jj05}), $i.e.$ after
the G-O rule has been violated by
  Cycle pair (24, 25).
 \inlinecite{jj15} predicted a violation of the G-O rule 
 Cycle pair (24, 25),  
 and  here  the results from the  
 analysis of two extended datasets strongly support it.
 We have 
extensively extracted the necessary information from   the trends
in the yearly and the cycle-to-cycle variations in 
$R_{\rm Z}$ ($i.e.$, mainly used the epochs of the violations of the G-O and
 inverse G-O rules).

\acknowledgements{The author thanks 
the anonymous referee for the critical review and useful comments 
and suggestions, and  Matthew Owens, Associate Editor, for useful 
editorial comments and suggestions. 
The author is thankful to  Ferenc V\'aradi for providing 
 the Sun's orbital angular momentum  data used here.}

Disclosure of Potential Conflicts of Interest: The authors declare that they
 have no conflicts of interest.

{}
\end{article}
\end{document}